\documentclass[12pt]{article}
\usepackage{a4}\usepackage{epsf}
\newcommand{\NO}{Nielsen-Olesen }
\newcommand{\uv}{ultraviolet }
\newcommand{\be}{\begin{equation}}
\newcommand{\ee}{\end{equation}}
\newcommand{\bea}{\begin{eqnarray}}
\newcommand{\eea}{\end{eqnarray}}
\newcommand{\la}{\lambda}
\newcommand{\nn}{\nonumber}
\newcommand{\pa}{\partial}
\newcommand{\Tr}{{\rm Tr~}}
\newcommand{\Ref}[1]{(\ref{#1})}
\newcommand{\E}{{\cal E}}
\renewcommand{\P}{{\cal P}}

\begin{document}
\title{Fermionic Vacuum Energy from a Nielsen-Olesen Vortex} 
\author{
{\sc M. Bordag}\thanks{e-mail: Michael.Bordag@itp.uni-leipzig.de} \\
{\small and}\\
I. Drozdov\thanks{e-mail: Igor.Drosdow@itp.uni-leipzig.de}\\
\small  University of Leipzig, Institute for Theoretical Physics\\
\small  Augustusplatz 10/11, 04109 Leipzig, Germany}
\maketitle
\begin{abstract}
  We calculate the vacuum energy of a spinor field in the background of
  a Nielsen-Olesen vortex. We use the method of representing the
  vacuum energy in terms of the Jost function on the imaginary
  momentum axis. Renormalization is carried out using the heat kernel
  expansion and zeta functional regularization. With this method well
  convergent sums and integrals emerge which allow for an efficient
  numerical calculation of the vacuum energy in the given case where
  the background is not known analytically but only numerically.  The
  vacuum energy is calculated for several choices of the parameters
  and it turns out to give  small corrections to the classical
  energy.
\end{abstract}
%%%%%%%%%%%%%%%%%%%%%%%%%%%%%%%%%%%%%%%%%%%%%%%%%%%%%%%%%%%%%%%%%%%%%%%%%%%%%%%
\section{Introduction}\label{Sec1}
%%%%%%%%%%%%%%%%%%%%%%%%%%%%%%%%%%%%%%%%%%%%%%%%%%%%%%%%%%%%%%%%%%%%%%%%%%%%%%%
Quantum corrections to classical background configurations are a topic
of continuing interest.  At present it is stimulated by the observation
made in lattice calculations that the field configurations responsible
for confinement are dominated by monopole or string like
configurations.  Another motivation comes from the stability analysis
of Z-strings with respect to fermionic fluctuations.

During the last years quantum corrections to string like
configurations have been investigated quite actively.  In
\cite{Bordag:1998tg} for the background of a finite radius magnetic
flux tube in QED the vacuum energy of a spinor was calculated. There
the method of representing the vacuum energy in terms of the Jost
function of the related scattering problem taken on the imaginary
momentum axis was applied.  This method has been developed earlier in
\cite{Bordag:1996fv} for spherically symmetric scalar background
fields.  The specific example considered in \cite{Bordag:1998tg} was a
homogenous magnetic field inside the flux tube. This investigation was
extended to more complicated profiles of the magnetic field inside the
flux tube in \cite{Scandurra:2000wr,Drozdov:2002um}.  In
\cite{Dunne:1998kw} a magnetic background was considered which depends
only on one spatial coordinate. In addition, this dependence is of a
form that the corresponding wave equation has an explicit solution so
that the vacuum energy could be calculated quite easily.  A more
general approach to fermionic vacuum energy was taken in
\cite{Fry:2001jw} where general bounds on the fermionic determinant
were obtained. Another approach was taken in
\cite{Pasipoularides:2000gg} where several profiles of the magnetic
background were considered and compared with the derivative
expansion. In \cite{Pasipoularides:2003jt} the limit of a strong
magnetic field was investigated in more detail.

 There is still an interest in backgrounds of infinitely thin strings
which constitute a singular background.  Typical for these
configurations is the need to apply the method of self adjoint
extensions. The object to consider in such examples is the vacuum
energy density per unit volume rather than the 'global' vacuum energy
which is for a string the density per unit length. For recent
investigations see, for example, \cite{Langfeld:2002vy} and
\cite{Diakonov:1999gg}. The problem with infinitely thin strings is
that their classical energy is infinite. Also, there are additional
counter terms. In the language of heat kernel expansion there are
additional contributions to the heat kernel coefficients, for instance
coefficients with half integer number, which reside on the surface
where the singularity is located. Another related example was
considered in \cite{Scandurra:2000wr} where the background is a finite
radius flux tube with the magnetic field concentrated on the surface
of the string.  The advantage of singular (and non smooth like in
\cite{Bordag:1998tg}) backgrounds can be seen in the usually quite
explicite formulas for the quantum fluctuations.  So in
\cite{Langfeld:2002vy,Diakonov:1999gg,Scandurra:2000wr} only Bessel
functions appear and in \cite{Bordag:1998tg,Drozdov:2002um} hyper
geometric functions in addition.

In general, physical background configurations should have a finite
energy, hence strings should have a non zero radius. A typical example
is the \NO string \cite{Nielsen:1973cs}. But here, not only the vacuum
fluctuations have to be calculated numerically, but even the
background itself.  The problem appears to have a calculational scheme
which does not need explicite formulas and which allows for efficient
numerical evaluation. There the main problem comes from the \uv
divergencies.  In analytical terms it is well known how to handle
them. First one has to introduce some intermediate regularization.
After that one has to subtract the counter terms and finally to remove
the regularization resulting now in a finite result.  However,
consider the last step in zeta functional regularization. Here one has to
perform an analytical continuation. Or let us consider some cut-off
regularization, where one has to remove the contributions proportional to
non negative powers of the cut-off parameter.  In principle such a
procedure can be done numerically (there are some examples) but this
is quite complicated and ineffective.  It is better to transform the
expression for the vacuum energy in a way that the final removal of
the regularization can be performed analytically so that only well
convergent sums and integrals remain. Such a method had been developed
in \cite{Bordag:1995jz,Bordag:1996fv} and in
\cite{Bordag:1998tg,Bordag:2002sa} applied to strings of finite
radius. The method is based on a representation of the regularized
vacuum energy in terms of the Jost function of the related scattering
problem taken on the imaginary momentum axis. Another method, using
phase shifts and momenta on the real axis was used in
\cite{Graham:2002fi} (see also \cite{Graham:2002xq} and references
therein), mainly for spherically symmetric backgrounds. Also, using
phase shifts, the case of a color magnetic vortex was considered in
\cite{Diakonov:2002bx}.  In a similar way in \cite{Groves:1999ks} the
vacuum energy for an electroweak string had been considered, where,
however, a step profile was taken in the final stage.

A completely different method is worth to be mentioned. In
\cite{Gies:2001tj,Gies:2001zp} world line methods were applied to the
calculation of vacuum energy which have the advantage not to rely on
separation of variables and therefore to be applicable for much more
general background configurations.

In the present paper we calculate the vacuum energy of a fermion in
the background of a \NO string. We use the method of representing the
regularized vacuum energy in terms of the Jost function on the
imaginary momentum axis which was developed in
\cite{Bordag:1996fv}. For the \NO vortex the background potential is
given only as a numerical solution of the corresponding equations of
motion. We use zeta functional regularization and determine the
counter terms from standard heat kernel expansion. The renormalized
vacuum energy is divided into two parts, the 'finite' and the
asymptotic ones, by subtraction and addition of some first terms of
the uniform asymptotic expansion of the Jost function which is
obtained using the Lipmann-Schwinger equation. In the asymptotic part
the analytic continuation in the regularization parameter is performed
analytically and well convergent double integrals remain. The 'finite'
part is represented as a well convergent sum and integral involving
the Jost function which is calculated from the numerical solutions of
the corresponding wave equation. Using these tools, the dependence on
the parameters of the considered model is investigated numerically.

The paper is organized as follows. In the next section the basic
notations for the considered model are introduced, the renormalization
is discussed and the basic formulas for the representation of the
vacuum energy are given. In the third section the asymptotic part of
the Jost function is derived and the asymptotic part of the vacuum
energy as well. In the fourth section the 'finite' part of the vacuum
energy is derived and the convergence properties are
discussed. Sect. 5 contains the numerical part of the work and Sect. 6 the
conclusions. 

\noindent
We use units where $\hbar=c=1$.

%%%%%%%%%%%%%%%%%%%%%%%%%%%%%%%%%%%%%%%%%%%%%%%%%%%%%%%%%%%%%%%%%%%%%%%%%%%%%%%
\section{Basic notations}\label{Sec2}
%%%%%%%%%%%%%%%%%%%%%%%%%%%%%%%%%%%%%%%%%%%%%%%%%%%%%%%%%%%%%%%%%%%%%%%%%%%%%%%
The Abelian Higgs model contains a $U(1)$ gauge field, $A_\mu(x)$, and
a complex scalar field, $\Phi(x)$. The action is
\be\label{AH}S=\int d^4x \ \left(
-\frac14 F_{\mu\nu}^2
+ \mid D_\mu \Phi |^2-\la\left(|\Phi|^2-\frac{\eta^2}{2} \right)^2\right),
\ee
where $D_\mu=\pa_\mu-i q A_\mu$ is the covariant derivative and
$F_{\mu\nu}=\pa_\mu A_\nu-\pa_\nu A_\mu$ is the field strength, for
more details see \cite{Achucarro:1999it}. The vacuum configuration is
given by $A_\mu=0$, $\Phi=\eta e^{ic}/\sqrt{2}$ where $c$ is some
constant and $\eta$ is the Higgs condensate. This configuration has
zero energy. A configuration with finite non zero energy must be at
spatial infinity in the vacuum manifold. Hence asymptotically the
gauge potential must be a pure gauge and the scalar field must tend to
a constant times an angular dependent phase.  A configuration of such
type is the Nielsen-Olesen string \cite{Nielsen:1973cs}. In
cylindrical coordinates $(x,y,z)\to(r,\varphi,z)$ one makes the ansatz
\be\label{ansatz}\Phi=\frac{\eta}{\sqrt{2}} f(r) e^{i n \varphi}, \ \
\ q A_\varphi = n v(r),
\ee
where $A_\varphi$ is the angular component of the vector
potential. The profile functions satisfy the boundary conditions
\be\label{bc} f(0)=v(0)=0, \ \ \ \ \ \ f(r)\to 1, v(r)\to 1  \  \ {\rm
  as}  \  \ r\to\infty.
\ee
Here $n$ is the winding number and gives at once the magnetic flux in
the string. In the following we consider $n=1$ only. The equations of
motion imply
\bea\label{eom}f''(r)+\frac1r
f'(r)-\frac{1}{r^2}\left(1-v(r)\right)^2f(r)+
\la\eta^2\left(1-f(r)^2\right)f(r)&=& 0 ,
\nn \\
v''(r)-\frac1r v'(r)+q^2 \eta^2 f(r)^2\left(1-v(r)\right)&=& 0 .  
\eea
After a rescaling, $r=\rho/(q\eta)$, these equations depend in fact
only on the combination $\beta=2\la/q^2$, which is at once the squared
ratio of the Higgs and vector masses. For $\beta=1$ the system
exhibits an additional symmetry and can be reduced to two first order
equations (Bogomolny equations). For $\beta<1$ the system is stable
for all values of the flux $n=1,2,\dots$, for $\beta>1$ only $n=1$ is
stable (a vortex with $n>1$ decays into $n$ vortexes with $n=1$).

The system of equations \Ref{eom} together with the boundary
conditions \Ref{bc} does not have an analytical solution and one is
left with numerical methods. The simplest way is to set the
derivatives in zero, i.e. the constants $a$ and $b$ the expansion for
small $r$, $f(r)=a r +\dots$ and $v(r)=b r^2 +\dots$, and to numerically
integrate the equations from $r=0$ to larger values of $r$. One has to
adjust these constants in a way that for large $r$ the asymptotic
values $f=1$ and $v=1$ are approached. Examples are shown in Fig. 1
for two values of $\beta$ with $q\eta=1$. The corresponding values
of the derivatives are $(a,b)=(0.26817,0.17481)$ for $\beta=0.3$ and
$(a,b)=(0.79958,0.42848)$ for $\beta=6$. As the asymptotic values for
$r\to\infty$ are reached with exponential speed it is in fact
sufficient to consider $r$ up to $\approx 15$. 
\begin{figure}[t]  \label{fig1}
\epsfxsize=14cm
\epsffile{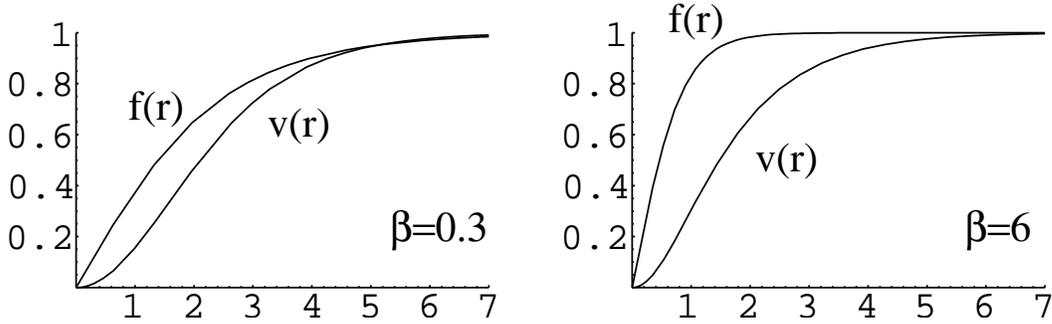}
% \end{picture}
\caption{The profile functions of the Nielsen-Olesen vortex as 
functions of the radius $r$ for $q\eta=1$.}
\end{figure}
sufficient to consider $r$ up to $\approx 15$. 

The classical energy (more exactly, the energy density per unit
length) of these configurations is given by
\bea\label{eclass}E_{\rm class}&=&\pi\int_0^\infty dr \ r \ 
\left(   \frac{1}{q^2 r^2}v'(\rho)^2
+\eta^2 \left(
f'(\rho)^2 
+\frac{1}{r^2}(1-v(\rho))^2 f(\rho)^2\right)
 \right. \nn \\   && \left. \qquad \qquad \qquad
+\frac{\la}{2} \eta^4 (1-f(\rho)^2)^2\right).
\eea
We calculate the background for $q\eta=1$ and restore later the
general setting by inserting $\rho=q\eta r$ in \Ref{eclass}.

The spinor is taken as a four component Dirac spinor with a
coupling to the background given by the  Lagrange density
\be\label{Lspinor}L=-i \overline{\Psi} \left( i\gamma^\mu D_\mu-f_e\mid
\Phi \mid \right) \Psi .
\ee
This model is chosen for the reason of simplicity. It provides a
coupling of the spinor not only to the vector but also to the scalar
background and it is motivated by the Yukawa coupling in the fermionic
sector of the standard model. We note that the interaction in
\Ref{Lspinor} is gauge invariant and that the coupling constant $f_e$
is dimensionless.  However, this model has a drawback. Since there is
only one spinor we are forced to take the absolute value,
$\mid\Phi\mid=\sqrt{(\Re \Phi)^2 +(\Im \Phi)^2}$, of the complex
scalar field (in the standard model there we don't  need to do so).  As a
consequence, the interaction is non polynomial. This is a
complication, for example if considering the vacuum energy together
with the dynamics of the background. Also, there is a need for an
additional counter term (see below).  However, because we are not
going to consider the dynamics of the background, this complication
does not affect the calculation of the vacuum energy if we let enter
the additional counter terms into the classical energy with a coefficient
which is put equal to zero after the renormalization is carried out. 
In addition, if we consider the same problem of calculating the vacuum
energy in the standard model then there is no such complication due to
its renormalizability.

The background is static and due to the translational invariance in
direction along the z-axis and the rotational invariance around the z-axis
($m=0,\pm1,\dots$ is the orbital angular momentum), the corresponding
momenta can be separated, $\Psi\to e^{-ip_0x^0+ip_3x^3}\left(\Psi_1
  e^{-(m+1)\varphi}\atop \Psi_2 \ \ e^{-m\varphi}\right)$.  After that
the Dirac equation decouples and one of the resulting  pair of two
component equations may be represented as
\be\label{spinoreq}\left(\begin{array}{ccc}
p_0-\mu(r) & , & \frac{\pa}{\pa r}-\frac{m-v(r)}{r} \\
-\frac{\pa}{\pa r}-\frac{m+1-v(r)}{r}& , &p_0+\mu(r)\end{array}\right)
\left(\begin{array}{c}\Psi_1\\ \Psi_2 \end{array}\right)=0.
\ee
The other pair is obtained by  changing the sign of $\mu$.  Here
the notation $\mu(r)=f_e\frac{\eta}{\sqrt{2}} f(r)$ has been
introduced. With respect to the spinor this is a radius dependent mass
density. From its value at infinity we define the spinor mass,
$m_e=f_e\frac{\eta}{\sqrt{2}}$. For a constant mass density the
problem is the same as for a pure magnetic flux tube which appears as
a special case in this way.

The vacuum energy of the spinor is given by the general formula
\be\label{Equant}\E_{\rm quant}=-\frac12 \sum_{(n)} e_{(n)}^{1-2s}.
\ee
Here $s$ is the zeta functional regularization parameter and $e_{(n)}$
are the one particle energies.  For simplicity we dropped the
arbitrary constant $\mu$ which is usually introduced to adjust the
dimension in zeta functional regularization. The quantum numbers $(n)$
include the sign of the one particle energies (all enter with positive
sign), the spin $s_z$ (two projections which we can account for by
summing over the sign of $\mu(r)$), the orbital angular quantum number
$m$ and the radial quantum number $n_r$ (assuming for a moment the
system being inserted into a large cylinder). In addition there is the
momentum $p_3$ which can be integrated according to
\[\label{p3}\int_{-\infty}^\infty \frac{dp_3}{2\pi} \
\left(p_3^2+x^2\right)^{\frac12-s} 
=
x^{2(1-s)}\frac{\Gamma(s-1)}{2\sqrt{\pi}\Gamma(s-\frac12)}
=
x^{2(1-s)}\frac{C_s}{4\pi s} 
\]
with $C_s=1+s(2\ln 2-1)+\dots$ and we arrive at
\be\label{Equant1}\E_{\rm quant}=-\frac{C_s}{4\pi s}  
\sum_{s_z, m, n_r} \left(e_{s_z, m, n_r}\right)^{2(1-s)}.
\ee
In order to get rid of the large cylinder we proceed as in
\cite{Bordag:1998tg}. We consider the cylindrical scattering problem
associated with Eq. \Ref{spinoreq} and define the Jost functions,
$f_m(k)$, with $p_0=\sqrt{m_e^2+k^2}$. We rewrite the sum over $n_r$
by an integral, tend the large cylinder to infinity and after dropping
the Minkowski space contribution we end up with
\be\label{Equant2}\E_{\rm quant}=\frac{C_s}{4\pi} \sum_{sgn
\mu}\sum_{m=-\infty}^\infty
\int_{m_e}^\infty dk \ \left(k^2-m_e^2\right)^{1-s} \ \frac{\pa}{\pa
  k} \ln f_m(i k)
\ee
which is our expression for the regularized ground state energy. Here
the integration is turned to the imaginary axis, more specifically, it
goes along the cut resulting from $\left(k^2-m_e^2\right)^{1-s}$. In
order to fully exploit the symmetry we renumber the orbital momenta by
$\nu$ according to
\be\label{nu}m=\left\{\begin{array}{rr} \nu-\frac12 &  (m\ge 0) \\[3pt]
                      -\nu-\frac12  & (m<0) \end{array}\right.
\ee
with $\nu=\frac12,\frac32,\dots$ and two signs of the orbital momentum
are to be taken into account. After rotation $k\to ik$ we define
$p=\sqrt{k^2-m_e^2}$ and for positive orbital momentum
Eq. \Ref{spinoreq} can be rewritten in the form
\be\label{speq}\left(\begin{array}{ccc}
ip-\mu(r) & , & \frac{\pa}{\pa r}-\frac{\nu-1/2-v(r)}{r} \\
-\frac{\pa}{\pa r}-\frac{\nu+1/2-v(r)}{r}& , &ip+\mu(r)\end{array}\right)
\left(\begin{array}{c}\Psi_1\\ \Psi_2 \end{array}\right)=0.
\ee
It can be seen (see below, Eqs. \Ref{eqg+} and \Ref{eqg-}) that for
the Jost function a change in the sign of $\mu$ corresponds to complex
conjugation and a change in the sign of the magnetic background,
$v(r)\to-v(r)$, corresponds to an exchange of positive and negative
orbital momenta. As a consequence, in Eq. \Ref{Equant2} we have a
factor of 4 and we take the real part of the half sum of positive and
negative orbital momenta.

The next step is the renormalization. We use the standard heat kernel 
expansion according to which the divergent part of the ground state
energy is given by (see, e.g., \cite{Bordag:1998tg} and we drop $a_0$)
\be\label{Ediv}E^{\rm   div}=
\frac{m_e^2}{32\pi^2}\left(\frac1s+\ln\frac{4}{m_e^2}-1\right) a_1
-\frac{1}{32\pi^2}\left(\frac1s+\ln\frac{4}{m_e^2}-2\right)a_2 ,
\ee
where $a_i$ are the standard heat kernel coefficients. We define the
renormalized ground state energy by $s\to0$
\be\label{Eren} E_{\rm   ren}=\E_{\rm quant}- E^{\rm   div}
\ee
in the limit $s\to0$. The definition \Ref{Eren} of the
renormalized ground state energy is chosen in a way that $E_{\rm ren}$
vanishes if the spinor mass $m_e$ taken as a parameter independent
from the background becomes large. This is the  so called 'large mass'
normalization condition. It has been discussed in detail in
\cite{Bordag:2000f,Bordag:1999vs} and in \cite{BGNV} it was shown to
be equivalent to the known 'no tadpole' condition.

The heat kernel
coefficients can be calculated using known methods from the squared
Dirac operator,
\bea\label{Dsquared}
\left(iD\hspace{-8pt}/+\mu(r)\right)\left(iD\hspace{-8pt}/-\mu(r)\right)&=&
-D^2+\frac{q}{2}F_{\mu\nu}\sigma^{\mu\nu}
-if_e \gamma^\mu \frac{\pa}{\pa x^\mu}\mid\Phi\mid
-f_e^2\mid\Phi\mid^2 \nn \\
&\equiv&-D^2+V
\eea
with $\sigma^{\mu\nu}=\frac{i}{2}[\gamma^\mu,\gamma^\nu]$.
The general expressions for the relevant heat kernel coefficients are
\bea\label{allghkks}a_1&=&\Tr \int d^2x \ (-V),  \nn \\
a_2&=&\Tr \int d^2x \left( 
-\frac{1}{12}F_{\mu\nu}^2+\frac12 V^2-\frac16 \Delta V \right),
\eea
where the trace is over the gamma matrices. Inserting for $V$ the
result is \bea\label{hkks}a_1 &=& -8\pi\int\limits_0^\infty dr \ r \
\left(\mu(r)^2-m_e^2\right), \nn \\ a_2 &=& 8\pi\int\limits_0^\infty
dr \ r \ \left(\frac13 \frac{v'(r)^2}{r^2} + \frac12 \left(
\mu'(r)^2+(\mu(r)^2-m_e^2)^2\right) \right)  \eea
(note that we consider densities per unit length of the string).
From Eq. \Ref{Ediv} we obtain the divergent part of the vacuum energy
in the form
\bea\label{Ediv2}\E^{\rm div}&=&-\frac{1-2s(\ln(m_e/2)+1)}{4\pi s}
\int\limits_0^\infty dr \ r \ 
\left(\frac13
\frac{v'(r)^2}{r^2}+\frac{f_e^2\eta^2}{4}f'(r)^2  \right. \nn \\
&& \left.   +\frac{f_e^4\eta^4}{8}\left(f(r)^4-1\right)\right)
-\frac{1}{4\pi}\int\limits_0^\infty dr \ r \ \frac{f_e^4\eta^4}{4} 
  \left(f(r)^2-1\right)     +O(s).
\eea
The interpretation is as follows. From the $v'(r)^2$-term we have the
standard renormalization of the electric charge $q$ in the classical
action \Ref{eclass}.  A renormalization of the scalar coupling
$\lambda$ absorbs the divergence proportional to $f(r)^4$.  After that
the remaining freedom is in a change of the condensate $\eta$.
However, this is obviously insufficient to absorb the remaining parts
of $\E^{\rm div}$.  In this way, in the classical energy an additional
structure must be present.  As suggested from Eq. \Ref{Dsquared}, it
is necessary to introduce at last a term proportional to
$\left(\frac{\pa}{\pa x_\mu}\mid\Phi \mid\right)^2$ into the action
\Ref{AH} and into the classical energy \Ref{eclass} as well.  It
should be noted that such a term is gauge invariant and that it has
the correct dimension. However, it represents another non polynomial
interaction. This is not surprising as the model given by
Eq. \Ref{Lspinor} itself contains a non polynomial interaction.
Accepting this we can finish the renormalization if we put the
coefficient in front of this term equal to zero after performing the
renormalization.

In order to perform the limit $s\to0$ in Eq. \Ref{Eren} we need to
rewrite the regularized ground state energy. For this we define the
asymptotic Jost function as the part of its uniform asymptotic
expansion for $\nu\to\infty$, $z\equiv\frac{k}{\nu}$ fixed, which
includes all powers up to $\nu^{-3}$,
\be\label{lnfas} \ln f_\nu(ik)= \ln f^{\rm as}_\nu
+O\left(\frac{1}{\nu^4}\right). 
\ee
Using $\ln f^{\rm as}_\nu$ we divide the energy into 
\be\label{Eren2}E_{\rm   ren}=\E_{\rm f}+ E^{\rm   as}
\ee
with the 'finite' part,
\be\label{Ef}\E^{\rm f}= \frac{1}{\pi}\sum_{\nu=\frac12,\frac32,\dots}
\int_{m_e}^\infty dk \ \left(k^2-m_e^2\right)  \ \frac{\pa}{\pa
  k} \left( \ln f_\nu(i k)-\ln f^{\rm as}_\nu(i k) \right)
\ee
and the 'asymptotic'  part,
\be\label{Eas}E^{\rm   as}=\frac{1}{\pi}\sum_{\nu=\frac12,\frac32,\dots}
\int_{m_e}^\infty dk \ \left(k^2-m_e^2\right)^{2(1-s)}  \ \frac{\pa}{\pa
  k}   \ln f^{\rm as}_\nu(i k) \ - \ E^{\rm   div}.
\ee
The sum and the integral in $\E^{\rm f}$, Eq. \Ref{Ef}, are finite by
construction of the asymptotic Jost function. Therefore we could put
$s=0$ therein. In the asymptotic part, $E^{\rm as}$, Eq. \Ref{Eas}, the
analytic continuation in $s$ has to be still performed which will be
done in the next section.

%%%%%%%%%%%%%%%%%%%%%%%%%%%%%%%%%%%%%%%%%%%%%%%%%%%%%%%%%%%%%%%%%%%%%%%%%%%%%%%
\section{The asymptotic part of the vacuum energy}\label{Sec3}
%%%%%%%%%%%%%%%%%%%%%%%%%%%%%%%%%%%%%%%%%%%%%%%%%%%%%%%%%%%%%%%%%%%%%%%%%%%%%%%
The asymptotic part of the Jost function can be derived from the
Lipmann-Schwinger equation just generalizing the procedure developed
in \cite{Bordag:1998tg}. The operator $\Delta\P$ there in Eq. (28)
reads now
\be\label{dP}\Delta\P=\left(
\begin{array}{ccc}\mu(r)&,&-\frac{v(r)}{r}\\-\frac{v(r)}{r}&,&-\mu(r)
\end{array}
\right).
\ee
Again, we have to perform iterations up to the fourth order in the
operator $\Delta\P$. Using the formulas given in
\cite{Bordag:1998tg} one arrives at the representation
\be\label{lnfas1}\ln f_{\nu}^{\rm as}(ik)=\int\limits_0^\infty 
\frac{dr}{r}\sum_{n=1}^3\sum_{j=n}^{3n}X_{nj}\frac{t^j}{\nu^n}
\ee
with $t=\frac{1}{\sqrt{1+(\nu z)^2}}$.  The coefficents are
\bea\label{Xnj}X_{11}&=&\frac12 v(r)^2+\frac12 r^2(\mu(r)^2-m_e^2),\nn \\
X_{13}&=&-\frac12 v(r)^2,\nn \\
X_{33}&=&\frac14  v(r)^2-\frac18 r^2 v'(r)^2-\frac18 v(r)^4-\frac14
r^2 v(r)^2(\mu^2-m_e^2)\nn \\
&&-\frac18 r^4 (\mu'(r)^2+(\mu^2-m_e^2)^2),\nn \\
X_{35}&=&-\frac{39}{16} v(r)^2+\frac18 r^2 v'(r)^2+\frac{3}{4} v(r)^4
+\frac34 r^2 v(r)^2(\mu^2-m_e^2)-\frac{3}{16}r^2 (\mu^2-m_e^2) ,\nn \\
X_{37}&=&\frac{35}{8}v(r)^2-\frac{5}{8} v(r)^4+\frac{5}{16}r^2
(\mu^2-m_e^2),\nn \\
X_{39}&=&-\frac{35}{16}v(r)^2.
\eea
The dependence on $v(r)$ is the same as in \cite{Bordag:1998tg} (where
it was denoted by $a(r)$), the dependence on $\mu(r)$ is new. In
Eq. \Ref{lnfas1} it had been integrated by parts in order to get the
shortest representation for the coefficients $X_{nj}$.

In $\E^{\rm as}$, Eq. \Ref{Eas}, the analytic continuation in $s$ can
be performed by rewriting the sum over the orbital momenta $\nu$ by
integrals using the Abel-Plana formula in the form as given in the
Appendix, Eq. \Ref{APla}.  From the first part, i.e., from the direct
integral over $\nu$, we get a contribution which just cancels $E^{\rm
div}$ in Eq. \Ref{Eas}. So we are left with the contribution from the
second part. Here we integrate over $k$ using the simple formula
\Ref{C3} and obtain
\be\label{Eas2}\E^{\rm as}=-\frac{C_s}{\pi} m_e^{2(1-s)}
\int_0^\infty\frac{dr}{r}\sum_{n,j} X_{nj}
\frac{\Gamma(2-s)\Gamma(s+\frac{j}{2}-1)}{\Gamma(j/2)} \ \Sigma_{nj}(rm_e)
\ee
with
\be\label{Signj}\Sigma_{nj}(x)=
\frac{1}{x^j}\int\limits_0^\infty\frac{d\nu}{1+e^{2\pi\nu}}
\frac1i \left(   
\frac{(i\nu)^{j-n}}{\left(1+\left(i\nu\over  x\right)^2\right)^{s+j/2-1}}-
\frac{(-i\nu)^{j-n}}{\left(1+\left(-i\nu\over  x\right)^2\right)^{s+j/2-1}}
\right).
\ee
The difference in the right hand side results from the deformation of
the integration contour in the Abel-Plana formula which gets tight
around the cut. Therefore the integration starts effectively from
$\nu=x$. 
This formula provides the best representation to perform the analytic
continuation in $s$. For $(n,j)=(1,1)$ we simply note
\be\label{Sig11}\Sigma_{11}(x)=\frac{2}{x^2}\int_x^\infty d\nu \ 
\frac{1}{1+e^{2\pi \nu}} \sqrt{\nu^2-x^2}.
\ee
Also for $(n,j)=(1,3)$ we can put $s=0$ immediately. However, we
prefer to integrate by parts in order to get a representation which is
in line with Eq. \Ref{Sig11}. In a similar way, integrating by parts,
we can proceed in all other contributions.  So the final form is
\be\label{Sigall}\Sigma_{nj}(x)=\frac{2}{x^2}\int_x^\infty d\nu \ f_{nj}(\nu)\
 \sqrt{\nu^2-x^2}
\ee
with
\[\label{fnj}
\begin{array}{rclrcl}
f_{11}(\nu)&=&\frac{1}{1+e^{2\pi \nu}}  , &
f_{13}(\nu)&=&-\left(\frac{\nu}{1+e^{2\pi \nu}}\right)' , \\[4pt]
f_{33}(\nu)&=&\left(\frac{1}{\nu\left(1+e^{2\pi \nu}\right)}\right)',
&
f_{35}(\nu)&=&
\left(\frac{1}{\nu}\left(\frac{\nu}{1+e^{2\pi \nu}}\right)'\right)' , \\[4pt]
f_{37}(\nu)&=&\frac13 
\left(\frac{1}{\nu}\left(\frac{1}{\nu}\left(\frac{\nu^3}{1+e^{2\pi
    \nu}}\right)'\right)'\right)' , &
f_{39}(\nu)&=&\frac1{15} 
\left(\frac{1}{\nu}\left(\frac{1}{\nu}\left(\frac{1}{\nu}\left(\frac{\nu^5}{1+e^{2\pi
    \nu}}\right)'\right)'\right)'\right)'.
\end{array}
\]
Using these formulas in Eq. \Ref{Eas2} we can put $s=0$ there and
insert for the coefficients $X_{nj}$ using Eqs.\Ref{Xnj}. After
rearranging contributions we arrive at
\bea\label{Eas3}\E^{\rm as}&=&\frac{-2}{\pi}\int_0^\infty\frac{dr}{r^3}
\left(v(r)^2 h_1(r m_e)+r^2v'(r)^2 h_2(r m_e)+v(r)^4 h_3(r
m_e)\right. \nn \\ && \left. \qquad +
r^2 v(r)^2 (\mu(r)^2-m_e^2) h_4(r m_e)+r^2  (\mu(r)^2-m_e^2) h_5(r
m_e)\right. \nn \\ && \left. \qquad+
r^4  (\mu'(r)^2 +(\mu(r)^2-m_e^2)^2) h_6(r m_e)   \right)
\eea
with 
\be\label{hj}h_j(x)=\int_x^\infty d\nu \ f_j(\nu) \ \sqrt{\nu^2-x^2}
\ee
and
\bea\label{fj}f_1(\nu)&=&-f_{11}(\nu)-f_{13}(\nu)+\frac12 f_{33}(\nu)
-\frac{13}{8}f_{35}(\nu)+\frac{7}{4}f_{37}(\nu)
-\frac{5}{8}f_{39}(\nu) , \nn \\ 
f_2(\nu)&=&-\frac14 f_{33}(\nu)+\frac1{12} f_{35}(\nu), \nn \\ 
f_3(\nu)&=&-\frac14 f_{33}(\nu)+\frac12 f_{35}(\nu)-\frac14
f_{37}(\nu), \nn \\ 
f_4(\nu)&=&-\frac12 f_{33}(\nu)+\frac12 f_{35}(\nu), \nn \\ 
f_5(\nu)&=&- f_{11}(\nu)-\frac18 f_{35}(\nu)+\frac18 f_{37}(\nu), \nn \\ 
f_6(\nu)&=&-\frac14 f_{33}(\nu).
\eea
According to these functions we divide the asymptotic part of the
ground state energy,
\be\label{Eas4}\E^{\rm as}\equiv \sum_{j=1}^6 \E^{\rm as}_j ,
\ee
into parts which have the meaning, for instance, of the asymptotic
part of the vacuum energy resulting from the magnetic background,
$\E^{\rm as}_1$, $\E^{\rm as}_2$ and $\E^{\rm as}_3$, or from the
mixed contribution, $\E^{\rm as}_4$, etc. We finish this section with
the remark, that the integrals in Eq. \Ref{Eas3} are all finite for
the considered background of a \NO vortex due to corresponding
properties of the functions $h_i(x)$, their behavior at $x\to0$ for
instance.   

%%%%%%%%%%%%%%%%%%%%%%%%%%%%%%%%%%%%%%%%%%%%%%%%%%%%%%%%%%%%%%%%%%%%%%%%%%%%%%%
\section{The 'finite' part of the vacuum energy and numerical results}
\label{Sec4}
%%%%%%%%%%%%%%%%%%%%%%%%%%%%%%%%%%%%%%%%%%%%%%%%%%%%%%%%%%%%%%%%%%%%%%%%%%%%%%%
In order to calculate the finite part of the vacum energy defined by
Eq. \Ref{Ef} we have, first of all, to set up a numerical scheme for
the calculation of the Jost function. The Jost function is defined as
the coefficients in the asymptotics for large radius in the so called
regular solution of the wave equation \Ref{spinoreq},
\be\label{seq1}\Psi^{(\pm)}(r)\sim   \frac12 \left\{ 
f_\nu^{(\pm)}(k) \Psi^{(\pm)}_{H^{(2)}}(kr)+
\overline{f}_\nu^{(\pm)}(k) \Psi^{(\pm)}_{H^{(1)}}(kr)    \right\},
\ee
where 
\be\label{seq1a}\Psi^{(\pm)}_{H^{(1,2)}}(kr)=\left(\begin{array}{c}
\sqrt{p_0+m_e} \ H^{(1,2)}_{\nu\pm(\frac12-\delta)}(kr) \\[6pt]
\pm\sqrt{p_0-m_e} \  H^{(1,2)}_{\nu\mp(\frac12+\delta)}(kr)
\end{array}\right)
\ee
($p_0=\sqrt{m_e^2+k^2}$) are the two linear independent solutions at
$r\to\infty$. The signs $(\pm)$ correspond to the sign of the orbital
momentum according to Eq. \Ref{nu} and $H^{(1,2)}_\mu(z)$ are Hankel
functions. In general, $\delta$ is the dimensionless value of the
magnetic flux, i.e. it is $v(r\to\infty)$, which is equal to one in
our case.

From Eq. \Ref{seq1}, the Jost function can be expressed in terms of the
solutions, 
\be\label{jost1}f_\nu^{(\pm)}(k)=\frac{\pi  r}{2i}\left(
\mp\sqrt{p_0-m_e} \psi_1(r) H^{(1)}_{\nu\mp(\frac12+\delta)}(kr) 
+\sqrt{p_0+m_e}  \psi_2(r) H^{(1)}_{\nu\pm(\frac12-\delta)}(kr)    \right),
\ee
where $\psi_{1,2}(kr)$ are the upper and lower components of
$\Psi(r)$.  Strictly speaking, as given by Eq. \Ref{jost1}, the
functions $f_\nu^{(\pm)}(k)$ depend on $r$ and only for $r\to\infty$
they tend to the Jost functions. However, for sufficiently large $r$
(larger than the scale of the background, say $r>15$ in the examples
shown in Fig. 1), Eq. \Ref{jost1} provides a good approximation. In
the considered case the background approaches its asymptotic values
with exponential speed. In the same way the difference between the
Jost function and the approximation \Ref{jost1} is small.

The regular solutions $\Psi(r)$ of Eq. \Ref{spinoreq} are defined as
becoming proportional to the free solutions for $r\to0$ and can be
expressed in terms of Bessel function,
\be\label{regsol}\Psi(r)\sim
\left(\frac{k}{q}\right)^{\nu+\frac12}
\left(\begin{array}{c}
 \sqrt{p_0+m_0} \ J_{\nu\pm\frac12}(qr) \\[6pt]
\pm \sqrt{p_0-m_0} \ J_{\nu\mp\frac12}(qr)
\end{array}\right)
\ee
with $q=\sqrt{k^2-m_e^2+m_0^2}$. We introduced for a moment $m_0=\mu(0)$
which corresponds to a more general case.  In all examples considered
below we will have $m_0=0$.  Note also the factor in front which
provides the correct normalization.

In order to actually carry out a numerical integration of
Eq. \Ref{spinoreq} it is useful to change to functions with regular
initial values at $r=0$. At once, we must change to imaginary
momenta. So we substitute $k\to ik$ and we make the ansatz
\bea\label{ans1}\Psi^{(+)}=i^\nu \left(\begin{array}{c}
 i\sqrt{p-im_0} \ \phi_1^{(+)} \\[6pt]
\sqrt{p+im_0} \ \phi_2^{(+)}
\end{array}\right) &,& 
\Psi^{(-)}=i^\nu \left(\begin{array}{c}
 \sqrt{p-im_0} \ \phi_1^{(+)} \\[6pt]
-i\sqrt{p+im_0} \ \phi_2^{(+)}
\end{array}\right).
\eea
The equations for $\phi^{(\pm)}$ are
\be\label{eq+}
\left(\begin{array}{ccc}
-(p+i\mu(r)) \ e^{i\theta} &,& \frac{\pa}{\pa r}-\frac{\nu-1/2-v(r)}{r}
\\
 \frac{\pa}{\pa r}+\frac{\nu+1/2-v(r)}{r}&,&-(p-i\mu(r)) \ e^{-i\theta}
\end{array}\right)\phi^{(+)} =0
\ee
and
\be\label{eq-}
\left(\begin{array}{ccc}
-(p+i\mu(r)) \ e^{i\theta} &,& \frac{\pa}{\pa r}+\frac{\nu+1/2+v(r)}{r}
\\
 \frac{\pa}{\pa r}-\frac{\nu-1/2+v(r)}{r}&,&-(p-i\mu(r)) \ e^{-i\theta}
\end{array}\right)\phi^{(-)} =0
\ee
with $e^{i\theta}=\sqrt{(p+im_e)/(p-im_0)}$.
Further we substitute
\be\label{ans+}\phi^{(+)}=\left(\frac{ q
  r}{2}\right)^{\nu-\frac12}\frac{1}{\Gamma(\nu+\frac12)}
\left(\begin{array}{r} \frac{qr}{2\nu+1} \ g_1^{(+)} \\ g_2^{(+)}
\end{array}\right)
\ee
and
\be\label{ans-}\phi^{(-)}=\left(\frac{ q
  r}{2}\right)^{\nu-\frac12}\frac{1}{\Gamma(\nu+\frac12)}
\left(\begin{array}{r}  g_1^{(-)} \\  \frac{qr}{2\nu+1} \ g_2^{(-)}
\end{array}\right).
\ee
The boundary conditions at $r=0$ for these functions are
\be\label{bcg}g^{(\pm)}_{(1,2)}(r)_{\mid_{r=0}}=1
\ee
and they obey the equations
\be\label{eqg+}\left(\begin{array}{ccc}
-(p+i\mu(r)) \frac{qr}{2\nu+1} \ e^{i\theta} &,& 
\frac{\pa}{\pa r}+\frac{v(r)}{r}
\\
 \frac{\pa}{\pa r}+\frac{2\nu+1-v(r)}{r}&,&
-(p-i\mu(r)) \frac{2\nu+1}{qr} \ e^{-i\theta}
\end{array}\right)\left(\begin{array}{c}g_1^{(+)} \\ 
g_2^{(+)} \end{array}\right)=0,
\ee
\be\label{eqg-}\left(\begin{array}{ccc}
-(p+i\mu(r))  \frac{2\nu+1}{qr} \ e^{i\theta}&,& 
\frac{\pa}{\pa r}+\frac{2\nu+1+v(r)}{r}
\\
 \frac{\pa}{\pa r}-\frac{v(r)}{r}&,&
-(p-i\mu(r)) \frac{qr}{2\nu+1} \ e^{-i\theta}
\end{array}\right)\left(\begin{array}{c}g_1^{(-)} \\ 
g_2^{(-)} \end{array}\right)=0.
\ee
The above mentioned symmetries can be seen in this representation
explicitely. The Jost functions read now
\be\label{fg+}
f_\nu^{(+)}(ik)=\frac{r(q r/2)^{\nu-\frac12}}{\Gamma(\nu+\frac12)}
\left(\frac{qr}{2\nu+1}wg_1^{(+)}(r) K_{\nu-\frac12-\delta}(kr)
+w^{*} g_2^{(+)}(r) K_{\nu+\frac12-\delta}(kr)\right)
\ee
and
\be\label{fg-}f_\nu^{(-)}(ik)=\frac{r(q r/2)^{\nu-\frac12}}{\Gamma(\nu+\frac12)}
 \left( wg_1^{(-)}(r) K_{\nu+\frac12+\delta}(kr)
+ \frac{qr}{2\nu+1} w^{*} g_2^{(-)}(r) K_{\nu-\frac12+\delta}(kr)\right)
\ee
with $w=\sqrt{(p+im_e)(p-im_0)}$ and modified Bessel
functions. Eqs. \Ref{eqg+} and \Ref{eqg-} together with the boundary
conditions \Ref{bcg} can be solved easily numerically. We used the
package {\verb NDSolve } in \verb Mathematica .

Next we need the asymptotic part of the Jost function. We use
Eq. \Ref{lnfas1} with the coefficients $X_{nj}$, Eq. \Ref{Xnj}. The
integrals over $r$  are convergent and can be calculated numerically
without any problems. Taking into account the mentioned symmetries we
drop all contributions which are odd under $v\to -v$ and we define the
subtracted logarithm of the Jost function as
\be\label{lnfsub}\ln f^{\rm sub}=
\frac12 \Re(\ln f^{(+)}_\nu(ik)+\ln f^{(-)}_\nu(ik))-\ln f_\nu^{\rm as}.
\ee
This expression we insert into $\E^{\rm f}$, Eq. \Ref{Ef}. After
integration by parts we arrive at
\be\label{Ef2}\E^{\rm f}=\sum_{\nu=\frac12,\frac32,\dots} \E^{\rm f}_\nu
\ee
with the contributions from the individual orbital momenta,
\be\label{Efnu}\E^{\rm f}_\nu=  \int_{m_e}^\infty
  dk  \ \  \E^{\rm f}_\nu (k),
\ee
and the integrands
\be\label{Efnuk}\E^{\rm f}_\nu (k)= -\frac{2}{\pi}  
   k \ \ln f^{\rm sub}.
\ee
As illustration of the convergence we show $\E^{\rm f}_\nu (k)$
enhanced by a factor of $k^3 \nu^3$ in Fig. 2 for several first $\nu$.
It is seen that all curves approach the same limiting value which is
essentially given by the next term in the uniform asymptotic expansion
of the logarithm of the Jost function after that included into $\ln
f^{\rm as}$, Eq. \Ref{lnfas} taking into account that the order
$\frac{1}{\nu^4}$ is absent for symmetry reasons and the order
$\frac{1}{\nu^5}$ is the next one.  In this way the integrals over $k$
are fast convergent.

In Fig. 3 we show the contributions from the individual orbital
momenta. The corresponding sum in Eq. \Ref{Ef2} is also fast
convergent and it is sufficient to take a quite small number of the
first orbital momenta.
\begin{figure}[t]  \label{fig2}
\epsfxsize=12cm 
\epsffile{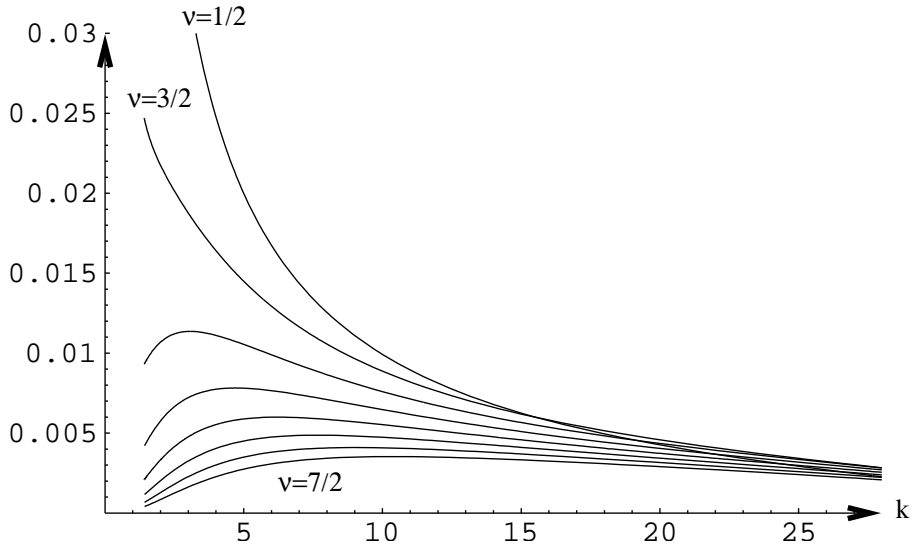}
\caption{The integrand $\E^{\rm f}_\nu (k)$ in Eq. \Ref{Efnu}
  multiplied by $k^3 \nu^3 $ as a function of $k$ for several first
  values of $\nu$. In this figure and in the next two the parameters
  are $\beta=0.3$, $q=0.5$, $f_e=1$, $\eta=1$.}
\end{figure}
\begin{figure}[t]  \label{fig3}
\epsfxsize=12cm 
\epsffile{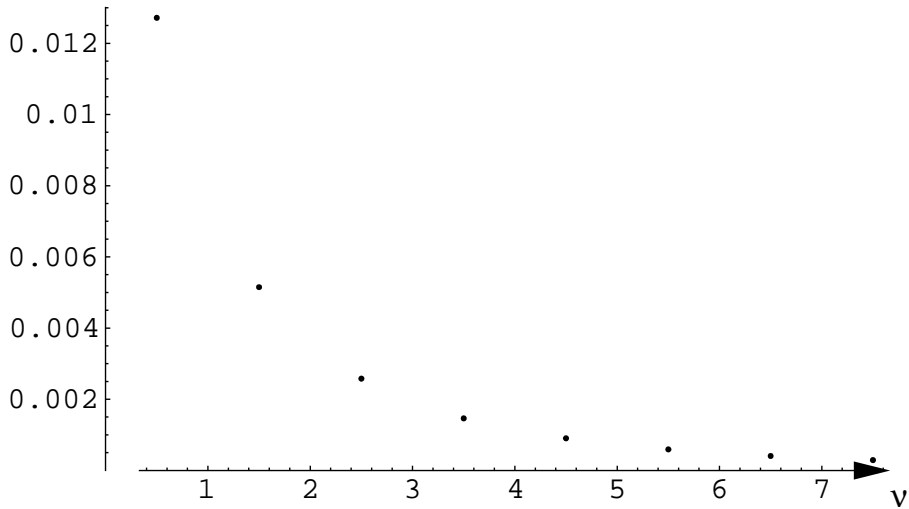}
\caption{The contributions from the individual orbital momenta
  $\E^{\rm f}_\nu$ to $\E^{\rm f}$ in Eq. \Ref{Ef2}.}
\end{figure}

We would like to add a note on bound states. They appear for imaginary
momenta, $k=i\kappa$, in Eq. \Ref{spinoreq}, in the region where
$p_0=\sqrt{m_e^2-\kappa^2}$ is real, $0<p_0<m_e$. For such momenta the
representation  \Ref{jost1} for the Jost function can be rewritten
in the form
\bea\label{jostbs}f^{(\pm)}_\nu(i\kappa)&=&-\lim_{r\to\infty}r \
i^{-\nu\mp(\frac12-\delta)}
\left\{ 
\sqrt{-p_0+m_e} \Psi_1(r) K_{\nu\mp(\frac12+\delta)}(\kappa r)\nn  
\right.\\ &&\left.
~~~~~~~~~~~~~~~
+\sqrt{ p_0+m_e} \Psi_2(r) K_{\nu\pm(\frac12+\delta)}(\kappa r) \right\}.
\eea
The functions $\Psi_{1,2}$ are solutions of the Eqs. \Ref{spinoreq}
with the initial conditions \Ref{regsol}. As both, equation and
initial conditions are real for the considered momenta the solutions
are also real. In this way the expression in the figure brackets in
Eq. \Ref{jostbs} are real. Their zeros just determine the location of
the bound states. As an example we plot in Fig. 4 these functions for
the two lowest orbital momenta. For orbital momentum $m=0$ we see one
bound state, for negative, $m=-1$, none. The appearance of the bound
states can be explained easily. Without scalar potential, i.e., for a
constant $\mu(r)=m_e$ the spinor moves in a pure magnetic field and in
the state where the magnetic moment is antiparallel to the magnetic
field its coupling just compensates the lowest Landau level resulting
in a zero mode known from \cite{AharonovCasherj}. Switching on the
scalar potential the spinor feels an additional attraction and becomes
a bound state. As concerns the vacuum energy these bound states are
outside of the integration region in Eq. \Ref{Equant2} and therefore
they do not complicate the calculations. As shown in
\cite{Bordag:1996fv} they are accounted for automatically in the
representation of the vacuum energy in terms of the Jost function. The
only we have to take care of is a possible appearance of bound states
on the imaginary $k$-axis in Eq. \Ref{Equant2} above $m_e$. This would
require a zero of the Jost functions in representation \Ref{fg+} or
\Ref{fg-} for real $k$ there. But these are genuine complex
expressions and it can be shown that they do not have such zeros.

\begin{figure}[t]  \label{figbs}
\epsfxsize=12cm 
\epsffile{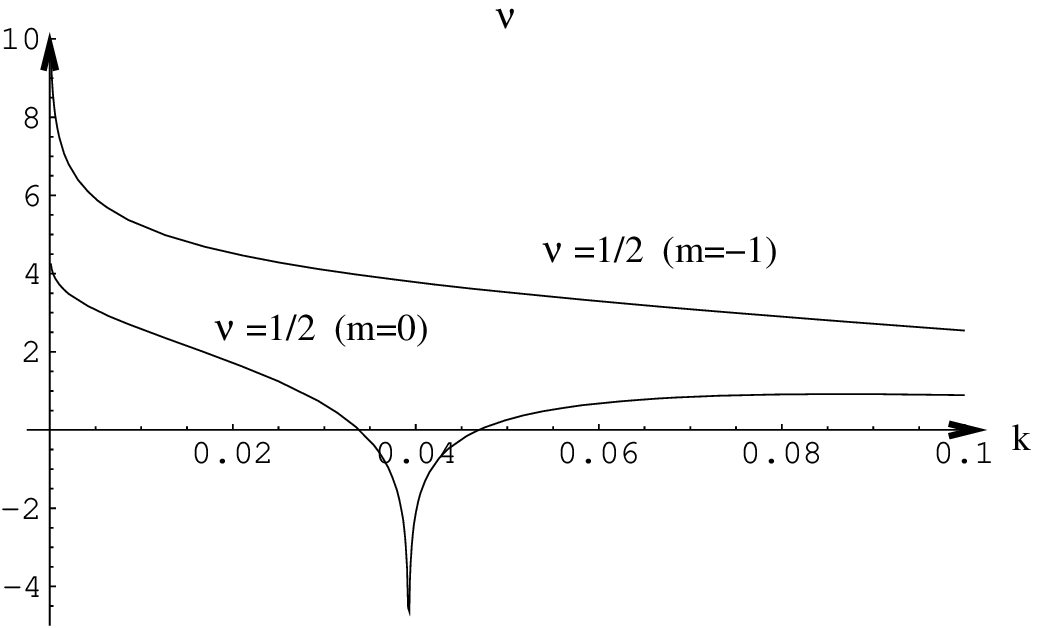}
\caption{The logarithm of the expressions in the figure brackets in
Eq. \Ref{jostbs}} as function of $\kappa$ for lowest orbital
momenta. The peak indicates a bound state. Its location is
$k=0.039232$.
\end{figure}

%%%%%%%%%%%%%%%%%%%%%%%%%%%%%%%%%%%%%%%%%%%%%%%%%%%%%%%%%%%%%%%%%%%%%%%%%%%%%%%
\section{Numerical results}\label{Sec5}
%%%%%%%%%%%%%%%%%%%%%%%%%%%%%%%%%%%%%%%%%%%%%%%%%%%%%%%%%%%%%%%%%%%%%%%%%%%%%%%
We investigated numerically the vacuum energy for values of  $\beta$
ranging from $\beta=0.3$ to $\beta=6$ for some choices of the remaining
parameters. The convergence properties of the sums and integrals
involved have been discussed in the previous section. Here, let us
first discuss the weight of the individual contributions. 

We start with the classical energy which is given by
Eq. \Ref{eclass}. It can be divided into three parts. The first one,
$E_{\rm mag}$, is the energy of the magnetic field. The second one,
$E_{\rm cov}$, results from the covariant derivative in Eq. \Ref{AH}
and is given by the second and the third contributions in
\Ref{eclass}. The third one, $E_{\rm scal}$, is the self energy of the
Higgs and is given by the fourth contributions in \Ref{eclass}. We
show in Table 1 two examples, one for smallest coupling of the scalar,
$\lambda=\beta q^2/2$ and the other for the largest value
considered. In both cases as well as in all intermediate ones, the
scalar self energy contribution is larger than the magnetic energy and
than $E_{\rm cov}$.

\begin{table}
\begin{tabular}{|lllll|}\hline
& $E_{\rm  mag}$  & $E_{\rm  cov}$ & $E_{\rm  scal}$ &$E_{\rm  class}$\\
  $\beta=0.3$ & 0.47913  & 1.50653 & 1.91653 & 3.90220\\
  $\beta=6$   & 0.97224  & 2.64606 & 3.88896  & 7.50726\\\hline
\end{tabular}
\caption{Two examples for the classical energy and its constituent parts for
 $q=0.5$, $\eta=1$.}
\end{table}

Next we consider for the parameters $q=0.5$, $f_e=1$, $\eta=1$ the
constituent parts of the asymptotic part of the vacuum energy, i.e.,
the $\E^{\rm as}_j$ as defined in Eq. \Ref{Eas4}. These quantities are
shown in Table 2. As it can be seen the contributions $\E^{\rm as}_5$
and $\E^{\rm as}_6$ are dominating. They result from the scalar
background, see Eq. \Ref{Eas3}.  But generally, all have at last a
numerical smallness of two orders of magnitude.

\begin{table}
\begin{tabular}{|lllll|}\hline 
&  $\E^{\rm as}_1 $ & $\E^{\rm as}_2 $ & $\E^{\rm as}_3 $ & $\E^{\rm as}_4$ \\
  $\beta=0.3$ & 0.00009139& -0.00006497& -5.5701 $10^{-8}$  &
  0.00001052\\
  $\beta=6$& 0.0004916& -0.0003590& -1.1546 $10^{-6}$  & 0.00004194
\\\hline 
 & $\E^{\rm as}_5 $ & $\E^{\rm as}_6 $ & $\E^{\rm as}=\sum_{j=1}^6 \E^{\rm as}_j $ &\\
$\beta=0.3$ &
  -0.004379&  -0.0017591 &-0.0061022 &\\
$\beta=6$&-0.003898&  -0.002483& -0.006208  &\\ \hline 
\end{tabular}
\caption{ the constituent parts of the asymptotic part of the vacuum
energy for $q=0.5$, $f_e=1$, $\eta=1$.}
\end{table}

Now, in Table 3, we represent the result of the calculations of the
complete vacuum energy for all considered examples. For two cases we
also represent the parts of the vacuum energy as a function of
$\beta$, see Figs. 5 and 6. From these results it can be seen that
there seems to be no general rule which part of the vacuum energy is
dominating and which not. Moreover, in dependence on the parameters
one part may be larger than other and smaller as well. It is even
impossible to say something definite about the sign of the vacuum
energy, it may change although in most cases it is negative.
 
\begin{table}
\begin{tabular}{|llllllll|}\hline 
$q$ &$\beta$& $f_e$&$\eta$& $E_{\rm  class}$& $\E^{\rm as}_1 $ &
  $\E^{\rm f} $ & $\E_{\rm ren}=\E^{\rm as} +\E^{\rm f} $ \\\hline
 0.5 & 0.3 & 1& 1 & 3.902 & -0.006102 & 0.02410 & 0.01800\\ 
0.5 & 0.6 & 1& 1 & 4.549 & -0.006102 & 0.01590 & 0.009795\\ 
0.5 & 1. & 1& 1 & 5.099 & -0.006104 & 0.01153 & 0.005421\\ 
0.5 & 3.5 & 1& 1 & 6.710 & -0.006146 & 0.004224 & -0.001922\\ 
0.5 & 6. & 1& 1 & 7.507 & -0.006208 & 0.001884 & -0.004325\\  \hline
1. & 0.3 & 1& 1 & 2.465 & -0.005766 & 0.008686 & 0.002920\\ 
1. & 0.6 & 1& 1 & 2.829 & -0.005665 & 0.005322 & -0.0003436\\ 
1. & 1. & 1& 1 & 3.142 & -0.005594 & 0.003030 & -0.002565\\ 
1. & 3.5 & 1& 1 & 4.093 & -0.005593 & -0.002588 & -0.008181\\ 
1. & 6. & 1& 1 & 4.591 & -0.005792 & -0.005077 & -0.01087\\  \hline
2. & 0.3 & 1& 1 & 2.105 & -0.003170 & -0.001883 & -0.005053\\ 
2. & 0.6 & 1& 1 & 2.399 & -0.002891 & -0.007285 & -0.01018\\ 
2. & 1. & 1& 1 & 2.652 & -0.002924 & -0.01182 & -0.01475\\ 
2. & 3.5 & 1& 1 & 2.652 & -0.002924 & -0.02451 & -0.02743\\ 
2. & 6. & 1& 1 & 3.861 & -0.006539 & -0.03022 & -0.03676\\  \hline
0.5 & 0.3 & 0.1& 0.1 & 0.03902 & -0.00006102 &-9.589 $10^{-6}$ &-0.00007061\\ 
0.5 & 0.6 & 0.1& 0.1 & 0.04549 & -0.00006102 & -0.00001214 & -0.00007316\\ 
0.5 & 1. & 0.1& 0.1 & 0.05099 & -0.00006104 & -0.00001421 & -0.00007525\\ 
0.5 & 3.5 & 0.1& 0.1 & 0.06710 & -0.00006146 & -0.00001975 & -0.00008121\\ 
0.5 & 6. & 0.1& 0.1 & 0.07507 & -0.00006208 & -0.00002216 & -0.00008425\\  \hline
0.5 & 0.3 & 0.1& 1 & 3.902 & -0.0003003 & -0.0009589 & -0.001259\\ 
0.5 & 0.6 & 0.1& 1 & 4.549 & -0.0005062 & -0.001214 & -0.001720\\ 
0.5 & 1. & 0.1& 1 & 5.099 & -0.0007110 & -0.001421 & -0.002132\\ 
0.5 & 3.5 & 0.1& 1 & 6.710 & -0.001423 & -0.001975 & -0.003398\\ 
0.5 & 6. & 0.1& 1 & 7.507 & -0.001814 & -0.002216 & -0.004030\\ \hline
\end{tabular}
\caption{The constituent parts of   of the vacuum
energy for all considered examples.}
\end{table}

\begin{figure}[t]  \label{fig5}
\epsfxsize=12cm 
\epsffile{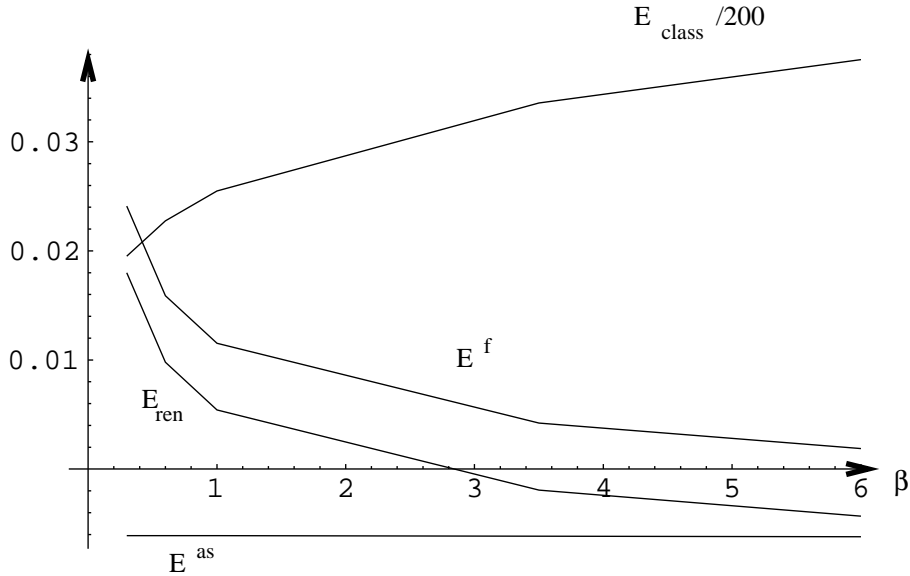}
\caption{The vacuum energy as a function of $\beta$ for the $q=0.5$,
  $f_e=1$, $\eta=1$. In order to represent all quantities within one
  plot the classical energy is divided by 200.} 
\end{figure}
\begin{figure}[t]  \label{fig6}
\epsfxsize=12cm 
\epsffile{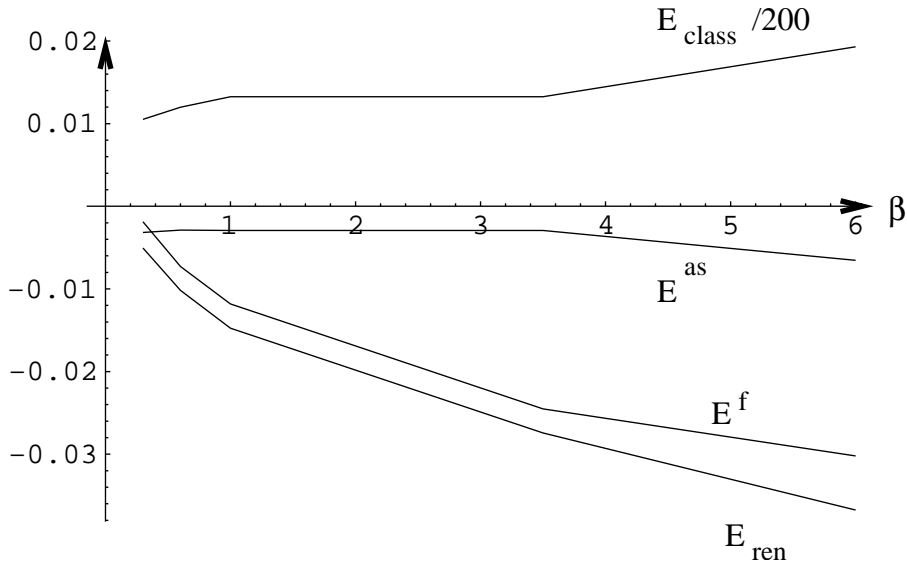}
\caption{The same as in figure 5 but for $q=2$.} 
\end{figure}

%%%%%%%%%%%%%%%%%%%%%%%%%%%%%%%%%%%%%%%%%%%%%%%%%%%%%%%%%%%%%%%%%%%%%%%%%%%%%%%
\section{Conclusions}\label{Sec6}
%%%%%%%%%%%%%%%%%%%%%%%%%%%%%%%%%%%%%%%%%%%%%%%%%%%%%%%%%%%%%%%%%%%%%%%%%%%%%%%
In the present paper we calculated numerically the vacuum energy of a
fermion in the background of a \NO string. As for renormalization we
used standard zeta functional regularization and determined the
counter terms from the first heat kernel coefficients (up to $a_2$).
It turned out that from the renormalization a counter term appears
which is not present in the initial action. It is gauge invariant and
it has the correct dimension but it represents a non polynomial
interaction. This has to be considered together with the non polynomial
interaction present in the model itself as discussed in Sect. 2.

The numerical investigations have been performed using methods
developed in the papers \cite{Bordag:1996fv,Bordag:1998tg}. In the
present paper the background is given purely numerically in difference
to the previous papers where it had been given analytically, for
example as a step function or some Gaussian profile. It has been
demonstrated that the computational scheme used here is well suited to
work with such backgrounds. This is a step forward to physically
really interesting problems. In the considered model the stability of
the background is given by topological arguments and for a realistic
choice of the parameters the quantum contribution is small. This
smallness has two sources. The one is the smallness of the coupling
constants which appears in front of the quantum contribution relative
to the classical one. The second one is a purely numerical smallness
of about two orders of magnitude as discussed in Sect. 5. It is
present even if the parameters and couplings are all of order one. It
is obviously connected with the dimension and the \uv renormalization.
So in \cite{Bordag:1995jz} it was found for a one dimensional
background that in (1+1) dimension the vacuum energy is by one order
of magnitude larger than for the same problem in (3+1)
dimensions. Comparing formulas (29) and (32) in \cite{Bordag:1995jz}
the interpretation is suggested that this difference is due to the one
additional \uv subtraction in the (3+1) dimensional case.

As seen in Sect. 5, for the dependence of the vacuum energy on the
various parameters no general rule can be seen so far. Even the sign
of the vacuum energy changes. The same applies to the relative weight
of the individual contributions. So sometimes the asymptotic part is
dominating, in other cases, however, the 'finite' part is larger. In
general, they are of the same order. From this one can conclude only
that in the given background there is no small parameter which could
allow for some approximative scheme. So for instance, if the
asymptotic part is dominating one could hope to get a good
approximation by including higher orders of the uniform asymptotic
expansion of the Jost function into the asymptotic part of the vacuum
energy and neglect the 'finite' part of it which is the numerically
much harder part of the problem. The examples in \cite{Bordag:2002sa}
have been of a kind suggesting this way in contrast to the example in the
present paper.

For the considered model of a spinor in the background of the \NO
vortex, due to its smallness, the vacuum energy has only a very small
influence on the dynamics of the background. Hence in the considered
case the vacuum energy is of limited physical importance. However, its
calculation gave new insights into the structure of such calculations
and demonstrated the power of the methods used.  A next step could be
to apply them to the $Z$ and electroweak strings where the stability
is not guaranteed by topological arguments and where the stability
issue is not finally settled with respect to the fermion contributions
\cite{Achucarro:1999it,Groves:1999ks}.
%%%%%%%%%%%%%%%%%%%%%%%%%%%%%%%%%%%%%%%%%%%%%%%%%%%%%%%%%%%%%%%%%%%%%%%%%%%%%%%
\section*{Acknowledgments} 
The authors are deeply indebted to V. Skalozub for valuable
discussions especially in an early stage of the work. Further they
thank V. Nikolaev for discussions on the numerical calculation of the
background and D. Vassilevich for helpful discussions and
suggestions. \\ One of the authors (I.D.) thanks the Graduiertenkolleg
{\it Quantenfeldtheorie} at the University of Leipzig for support.
%%%%%%%%%%%%%%%%%%%%%%%%%%%%%%%%%%%%%%%%%%%%%%%%%%%%%%%%%%%%%%%%%%%%%%%%%%%%%%%
%%%%%%%%%%%%%%%%%%%%%%%%%%%%%%%%%%%%%%%%%%%%%%%%%%%%%%%%%%%%%%%%%%%%%%%%%%%%%%%
\section*{Appendix}\label{App}
%%%%%%%%%%%%%%%%%%%%%%%%%%%%%%%%%%%%%%%%%%%%%%%%%%%%%%%%%%%%%%%%%%%%%%%%%%%%%%%
The Abel-Plana formula used in Sec.2 reads
\be\label{APla}
 \sum\limits_{\nu=\frac12,\frac32,\dots}f(\nu)=
\int\limits_{0}^{\infty}d \nu~f(\nu)
+\int\limits_{0}^{\infty}{d \nu\over 1+e^{2\pi\nu}}{f(i\nu)-f(-i\nu)\over i}.
\ee
The following formulas are used in the text,
\be\label{C3}
\int\limits_m^\infty dk(k^2-m^2)^{1-s}{\partial\over\partial  k}t^{j}
=
- m^{2-2s}
{\Gamma(2-s)\Gamma(s+\frac{j}{2}-1)\over  \Gamma(\frac{j}{2})}
 { \left({\nu\over mr }\right)^{j-n} \over
   \left(1+\left({\nu\over mr}\right)^{2}\right)^{s+\frac{j}{2}-1}}
\ee 
with $t=1/\sqrt{1+(kr/\nu)^2}$.  They can be easily derived, see also
(C3) and (C2) in \cite{Bordag:1998tg}.

%%%%%%%%%%%%%%%%%%%%%%%%%%%%%%%%%%%%%%%%%%%%%%%%%%%%%%%%%%%%%%%%%%%%%%%%%%%%%%%

\end{document}